\documentstyle[prl,aps,multicol,epsf]{revtex}

\begin{document}
\draft				% \draft command makes pacs numbers print
\preprint{ITP-UH-13/99}
\title{Doping Induced Magnetization Plateaus}
   
\author{Holger Frahm and Constantin Sobiella}
\address{Institut f\"ur Theoretische Physik, Universit\"at Hannover,
         D-30167~Hannover, Germany
}
\date{July 1999}
\maketitle
\begin{abstract}
The low temperature magnetization process of antiferromagnetic
spin-$S$ chains doped with mobile spin-$(S-1/2)$ carriers is studied
in an exactly solvable model.  For sufficiently high magnetic fields
the system is in a metallic phase with a finite gap for magnetic
excitations.  In this phase which exists for a large range of carrier
concentrations $x$ the zero temperature magnetization is determined by
$x$ alone.  This leads to plateaus in the magnetization curve at a
tunable fraction of the saturation magnetization.  The critical
behaviour at the edges of these plateaus is studied in detail.
\end{abstract}

%%%%%%%%%%%%%%%%%%%%%%%%%%%%%%%%%%%%%%%%%%%%%%%%%%%%%%%%%%%%%%%%%%%%%%
% body of paper here
\begin{multicols}{2}
Synthetization of new magnetic materials and availability of very high
magnetic fields provide new possibilities to study the magnetization
process of low-dimensional quantum spin systems.  In particular
so-called spin liquids realized in quasi-one dimensional
antiferromagnetic systems such as spin chains, spin ladders and
exchange-alternating spin chains attract much interest at present due
to the possible occurence of magnetization plateaus associated with
gapped excitations.  In addition to saturated magnetization $M_s$ such
plateaus, i.e.\ regions where the magnetization does not depend on the
magnetic field for sufficiently low temperatures, are admissible from
topological considerations at certain fractions of $M_s$ depending on
the value of the spin of the substance and the translational symmetry
of the ground state.  Necessary conditions for the occurence
of plateaus have been formulated by Oshikawa {\em et~al.}\ \cite{Oshi97}
employing a generalization of the Lieb-Schultz-Mattis theorem: for a
spin-$S$ chain with a magnetic unit cell containing $q$ magnetic moments
this feature can appear at rational values $\langle M\rangle$ with
integer $q(S-\langle M\rangle)$.
The existence of these phenomena in a variety of models has been
established by numerical and analytical studies of various
low-dimensional magnetic insulators including spin chains, spin
ladders and systems with multi spin exchange or exchange anisotropies
\cite{Hida94,Tone96,Okam96,Tots97,Cabra97,Cabra98,SaTa98,SaKi98,%
Fledd99,%
SaHa99,Momoi99,Chen99}.  
Very recently, several experimental observations of such magnetization
plateaus at non-zero $\langle M\rangle$ have been reported
\cite{Narumi98,Tana98,Shira98}.

A common feature in these systems is that the plateaus in the
magnetization curves appear at certain simple fractions of the
maximal value $M_s$ as a consequence of their topological origin.
In this letter we report on a mechanism leading
to gaps for magnetic excitations at magnetizations which can be
controlled by suitable preparation of the sample, namely doping.  We
study this phenomenon in the framework of a recently introduced class
of integrable models for doped Heisenberg chains which may be used as
a basis for studies of certain features of doped transition metal
oxides \cite{fpt:98,frahm:99}.  Starting from the double-exchange
model \cite{DE} a strong ferromagnetic Hund's rule
coupling between the spins of the itinerant $e_g$ electrons and
localized quantum spins $(S-{1}/{2})$ arising from the $t_{2g}$
electrons allows to introduce an effective Hamiltonian on a restricted
Hilbert space with maximally allowed spin $S'$ on a given lattice site
\cite{dagx:96,mhda:96}, i.e.\ $S'=S$ if the electronic state on this
site is occupied, or $S'=S-1/2$ if there is no $e_g$ electron (denoted
as a hole in the following).  This derivation of a low energy
Hamiltonian generalizes that of the $t$--$J$ model from the Hubbard
model \cite{zhri:88} which is corresponds to the case of $S=1/2$,
i.e.\ no localized spins.
Numerical studies of the $S\ge1$ variants of these models have been
performed to gain a better understanding of experimental findings for
the doped Haldane system Y$_{2-x}$Ca$_x$BaNiO$_5$ $(S=1)$
\cite{dagx:96} and manganese oxides such as La$_{1-x}$Ca$_x$MnO$_3$
($S=2$) \cite{dagx:98}.

Below we consider integrable models of this type in one spatial
dimension.  Similar to the models obtained from the general procedure
outlined above their Hamiltonians are of the form
\begin{equation}
   {\cal H}^{(S)} =\sum_{n=1}^L \left\{
        {\cal X}^{(S)}_{n,n+1}
        + {\cal T}^{(S)}_{n,n+1}\right\}\,.
\label{hamil1}
\end{equation}
Here ${\cal X}^{(S)}_{ij}$ and ${\cal T}^{(S)}_{ij}$ describe the
(antiferromagnetic) exchange and hopping of the holes between sites
$i$ and $j$ of the lattice, respectively.  $SU(2)$ invariance of
the model implies that they can be written as polynomials of the
operator ${\mathbf S}_i\cdot {\mathbf S}_j$ (see e.g.\
\onlinecite{mhda:96}).
The form of these polynomials is fixed in the integrable models
\cite{frahm:99}.  For example, in the $S=1$ case with possible relevance
to the doped Nickel oxides the antiferromagnetic exchange terms are given
in terms of bilinear and biquadratic Heisenberg couplings depending on
the values $S_{i,j}\in\left\{{1}/{2},1\right\}$ of the spins on sites $i$ and $j$
\cite{fpt:98}:
%\begin{equation}
\[
   {\cal X}^{(1)}_{ij} =
        {1\over2}\left(
         {1\over S_i S_j} {\mathbf S}_i \cdot {\mathbf S}_j - 1
         +\delta_{S_iS_j,1}\left(1-({\mathbf S}_i\cdot{\mathbf S}_j)^2\right)
        \right)\ .
\]
%\label{hex}
%\end{equation}
(Note that the undoped chain, i.e.\ $S_i=1$ for all $i$, is the integrable
spin-$1$ Takhtajan-Babujian model \cite{takh:82,babu:83} while the
completely doped chain is the spin-$1/2$ Heisenberg chain with bilinear
exchange.)
Similarly, the kinetic term of the integrable spin-$1$ model reads
%\begin{equation}
\[
   {\cal T}^{(1)}_{ij} =
        -\left(1-\delta_{S_i,S_j}\right)
         {\cal P}_{ij} \left( {\mathbf S}_i \cdot {\mathbf S}_j
	\right)\ ,
\]
%\label{hkin}
%\end{equation}
where ${\cal P}_{ij}$ is an operator permuting the states on sites $i$ and $j$
thereby allowing the spin-$1/2$ ``holes'' to propagate.  The additional
exchange term in this expression leads to different hopping amplitudes
$t(S_{ij})$ depending on the total spin $S_{ij}$ on the participating sites,
i.e.\ $t({1/2})=-1$ and $t({3/2})=+{1}/{2}$ for the possible values
$S_{ij}={1}/{2}$ and ${3}/{2}$, respectively.  These amplitudes differ from
the values proposed in Ref.~\onlinecite{dagx:96} for the doped Nickel oxide,
namely $t({1/2})=-1/2$, $t({3/2})=1$.  This is one reason for the absence of a
ferromagnetic phase in the integrable model (\ref{hamil1}) (see
\onlinecite{fpt:98} for a discussion of the other differences).

For general $S$ the integrable models are constructed from solutions of a
Yang-Baxter equation and can be solved by means of the algebraic Bethe Ansatz
\cite{frahm:99}.  Their thermodynamical properties at finite temperature $T$
can be obtained from the solution of the thermodynamic Bethe Ansatz (TBA)
equations, i.e.\ the following set of coupled nonlinear integral equations
\begin{eqnarray}
 && \epsilon_n(\xi) = Ts\ast\ln[1 + {\rm e}^{\epsilon_{n -1}(\xi)/T}]
                          [1 + {\rm e}^{\epsilon_{n +1}(\xi)/T}]
\nonumber\\
        && \qquad - 2\pi \delta_{n,2S}\ s(\xi)
           -\delta_{n,1}\ Ts\ast\ln[1 + {\rm e}^{- \kappa(\xi)/T}]\ ,
\nonumber\\
\label{eq:TBA}\\
  && -[ 2\pi a_{2S}\ast s(\xi) + \mu] 
     - Ts\ast\ln[1 + {\rm e}^{\epsilon_{1}(\xi)/T}]
\nonumber\\
  &&\qquad = \kappa(\xi) + TR\ast\ln[1 + {\rm e}^{- \kappa(\xi)/T}]\,.
\nonumber
\end{eqnarray}
Here $\left(f\ast g\right)(\xi)$ denotes a convolution in the space of
rapidities $\xi$, $a_n(\xi)=(2n/\pi) \left(4\xi^2+n^2\right)^{-1}$,
$s(\xi) =\left(2\cosh\pi\xi\right)^{-1}$ and $R = a_2\ast(1 +
a_2)^{-1}$.
Eqs.\ (\ref{eq:TBA}) are to be solved subject to the condition
$\lim_{n\to\infty}(\epsilon_n/n) = H$ with the external magnetic field $H$ and
$\mu$ is the chemical potential for the holes controlling their concentration.
In terms of the functions $\epsilon_n(\xi)$ and $\kappa(\xi)$ the free
energy of this system reads ($E_0^{(S)}$ is the ground state energy of the
spin-$S$ Takhtajan-Babujian chain for $H=0$ \cite{babu:83})
\begin{eqnarray}
 && \frac{1}{L}F(T,H,\mu) = \frac{E_0^{(S)}}{L}
	- T\int {\rm d}\xi\ s(\xi)
	  \ln[1 + {\rm e}^{\epsilon_{2S}(\xi)/T}]
\nonumber\\
      &&\qquad  - T\int {\rm d}\xi\ (a_{2S}\ast s)(\xi)
	\ln[1 + {\rm e}^{- \kappa(\xi)/T}]\ .
\label{FreeE}
\end{eqnarray}
The low temperature ($H\gg T$) phase diagram for the spin-1 system has
been obtained in Ref.~\onlinecite{fpt:98}, qualitatively the same
behaviour is found for general $S\ge1$ \cite{frahm:99}.

\begin{figure}
\begin{center}
\leavevmode
\epsfxsize=0.45\textwidth
\epsfbox{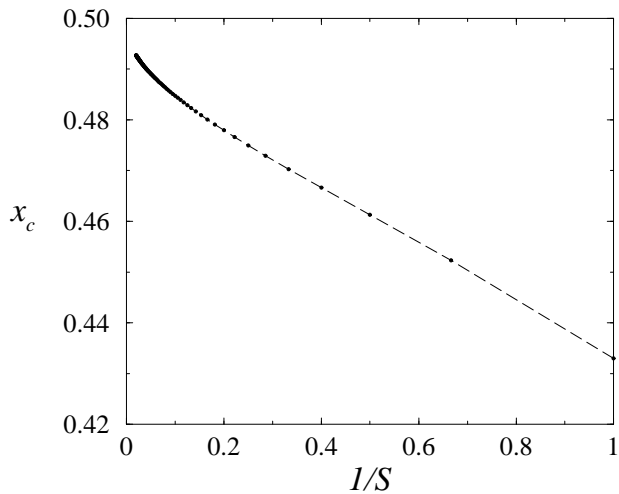}
\end{center}
\caption{Maximal concentration of holes $x_c(S)$ for the existence of
a magnetization plateau vs.\ $S^{-1}$ (the line is a guide to the eye only).
\label{fig:xmax}}
\end{figure}

\begin{figure}
\begin{center}
\leavevmode
\epsfxsize=0.45\textwidth
\epsfbox{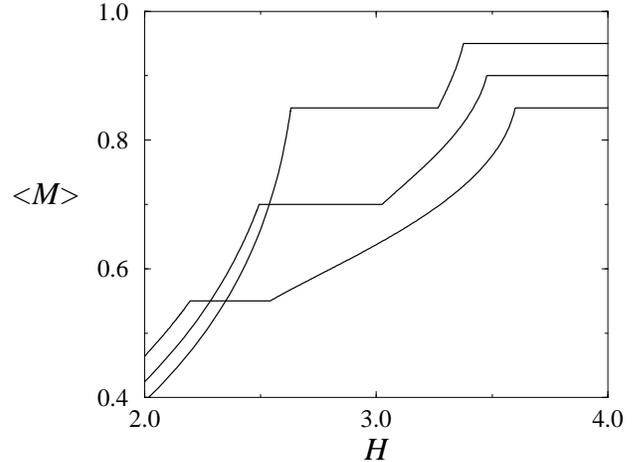}
\end{center}
\caption{Zero temperature magnetization curve of the doped $S=1$ chain
for  the hole concentrations $x=0.1$, $0.2$, $0.3$ (top to bottom).
\label{fig:plat10}}
\end{figure}

Here we study the properties of these systems in a magnetic field at
fixed doping.  For hole concentrations $0<x<x_c(S)$ (see Fig.\
\ref{fig:xmax}) the low energy excitation spectrum of the system
allows to identify four intermediate field phases (labelled A, B$_1$,
C and B$_2$ in Fig.~1 of Ref.~\onlinecite{fpt:98}) for $0<H<H_s$
before the system is completely polarized for $H>H_s$.  Of particular
interest in the present context is the phase C: since the system is
not ferromagnetically polarized one expects nontrivial excitations for
both the charge and magnetic degrees of freedom.  In the neighbouring
phases B$_{1,2}$ these excitations are massless leading to an
effective description of these phases in terms of a Tomonaga-Luttinger
model.  The analysis of the $T=0$ limit of the TBA eqs.\
(\ref{eq:TBA}) shows that in phase C only one of these modes is
gapless \cite{frahm:99}.  The resulting low-energy theory is that of a
single mode with dispersion
\begin{equation}
 \epsilon_{2S}(\xi) 
   + \int_{-Q}^Q {\rm d}\xi'\ K(\xi-\xi') 
	\epsilon_{2S}(\xi')
  =\epsilon_{2S}^{(0)}(\xi)
\label{eq:intC}
\end{equation}
where $K(\xi)=2\sum_{k=1}^{2S-1}a_{2k}(\xi)$, $\epsilon_{2S}^{(0)}(\xi)
=\left(2S-\frac{1}{2}\right)H-\mu -2\pi\sum_{k=1}^{2S} a_{2k-1}(\xi)$ and $Q$ is
a function of magnetic field and chemical potential through the condition
$\epsilon_{2S}(\pm Q)=0$.
The corresponding hole concentration $x=\int_{-Q}^Q {\rm d}\xi\
\sigma_{2S}(\xi)$ is obtained from an equation for $\sigma_{2S}$ similar to
(\ref{eq:intC}) with driving term replaced by $\sigma_{2S}^{(0)}(\xi) =
\sum_{k=1}^{2S} a_{2k-1}(\xi)$.
Further analysis of the zero temperature limit of the TBA eqs.\
(\ref{eq:TBA}) shows that the massless mode in this phase carries the
charge degrees of freedon while all magnetic excitations are gapped,
i.e.\ $\kappa(\xi)<0$ and $\epsilon_{n\ne2S}(\xi)>0$ for all
$\xi$.  
Hence, $x$ \emph{and} the magnetization $M_p=S-{3 x}/{2}$ are constant
throughout this phase for fixed $Q$, i.e.\ $\left(2S-\frac{1}{2}\right)
H-\mu ={\rm const}$.  This implies plateaus in the magnetization curve $M(H)$
\emph{below} the saturated value $M_s=S-x/2$ (see Fig.~\ref{fig:plat10} for
$S=1$).  The end points of these plateaus are $H_{c1} = -2\mu$ and
\begin{equation}
  H_{c2} = 2\mu +\frac{4}{S} + 2\int_{-Q}^Q {\rm d}\xi\
	a_{2S-1}(\xi)\epsilon_{2S}(\xi)\ .
\end{equation}
As $H\to H_{c1,2}$ from inside the plateau region the spin gap closes
as $\Delta \propto \left| H-H_{c1,2} \right|$.

\begin{figure}
\begin{center}
\leavevmode
\epsfxsize=0.4\textwidth
\epsfbox{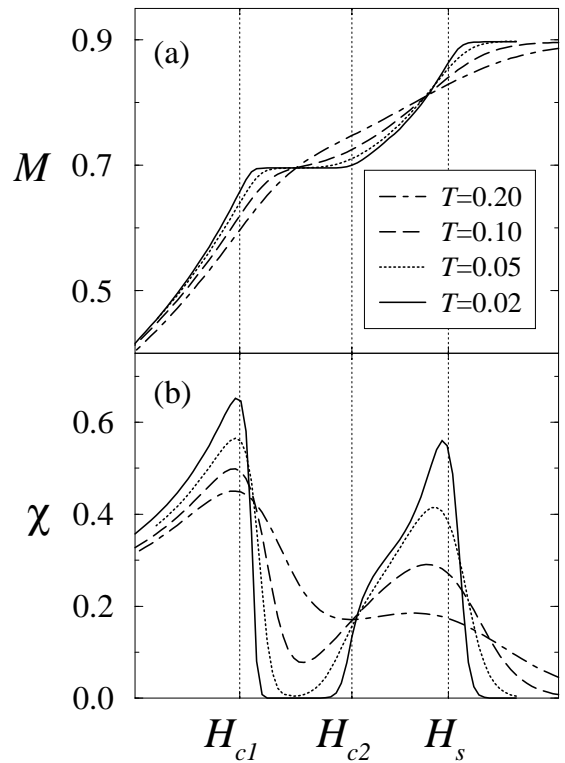}
\end{center}
\caption{Magnetization curve (a) and magnetic susceptibility (b) of
the doped $S=1$ chain with hole concentration $x=0.2$ at temperatures
$T=0.02$, $0.05$, $0.1$, $0.2$ (in units of (\protect{\ref{hamil1}})).
\label{fig:plat1T}}
\end{figure}

For finite temperatures the full set of TBA eqs.\ has to be solved to
determine the magnetization curves.  In a sufficiently strong magnetic
field $H\gg T$ however, the energies $\epsilon_{n>2S}$ are gapped and
can be eliminated from the TBA eqs.\ (\ref{eq:TBA}) \cite{fpt:98}.
For the doped $S=1$ chain this procedure leads to a coupled set of
three nonlinear integral equations which are straightforward to solve
by iteration.
Choosing the chemical potential such that the hole concentration
$x=-\partial F/\partial \mu$ is fixed the magnetization and
magnetic susceptibility can be obtained from the thermodynamical potential
$\Omega(T,H,x)=F(T,H,\mu)+\mu x$.  In Fig.~\ref{fig:plat1T} we present
the resulting data for various temperatures.  They clearly show the
formation of plateaus with decreasing temperature and the singular
behaviour arising in the vicinity of the transitions into the spin gap
phase as expected from the analysis of the zero temperature phase
diagram.

Remarkably, the nature of these singularities on the two critical end
points $H_{c1,2}$ of the plateau is quite different: for the magnetic
insulators discussed in the introduction the singular part of the
magnetization near the plateaus has been predicted to show a square
root behaviour \cite{Oshi97} due to the similarity of the transition
to a commensurate-incommensurate transition \cite{CIC}.
A reliable numerical verification of this prediction --- even
for an integrable model --- is extremely difficult for transitions other
than the one into the ferromagnetically polarized state \cite{Cabra98}.
In the model considered here such diffulties arise for $H\to H_{c1}$
only: for $H \lesssim H_{c1}$ the zero temperature magnetization
shows a critical behaviour $\propto \left({H_{c1} -H}\right)^\alpha$
consistent with the square root behaviour $\alpha=1/2$ within
the numerical accuracy of our data.
On the other hand, near $H_{c2}$ the magnetization depends linearly on
the external field, i.e.\ $M-M_p\propto H -H_{c2}$ for $H\gtrsim
H_{c2}$.  This difference in the critical behaviour is also evident in
the temperature dependence of the magnetic susceptibility near
$H_{c1,2}$, in particular $\chi\approx {\rm const.}$ at the high-field
end of the plateau (see Fig.~\ref{fig:plat1T}(b)).  Note, that a
similar $T$-dependence has been observed in experiments on certain
spin-$1/2$ Heisenberg ladders \cite{Chab97,Shira97}.

This different singular behaviour is a consequence of the coupling
between the two massless excitations present in the Tomonaga-Luttinger
phases outside the interval $H_{c1}<H<H_{c2}$: without the magnetic
field breaking the spin-$SU(2)$ these modes can be assigned usually
to spin and charge excitations separately based on their different
symmetries.  In an external field, however, this assignment leads to
a coupling of the two sectors.  In certain cases this interaction can
be removed by allowing for mixing of the corresponding quantum numbers
\cite{fn1}.
Analysing the Bethe Ansatz wave functions we can determine the
relation between the total charges $Q_{1,2}$ in the two gapless
sectors to the physical quantum numbers, i.e.\ number of holes and
$z$-component of the total spin.  In the phase for $H<H_{c1}$ a change
in magnetization at fixed doping affects the total charge in one of
the sectors only.  Necglecting the coupling to the other sector,
bosonization then gives the familiar square root singularity of
the magnetization at $H\lesssim H_{c1}$ \cite{CIC}.
For $H>H_{c2}$, however, a change of magnetization requires $\Delta
Q_1=-\Delta Q_2$ for fixed doping $x$.  As the critical field $H_{c2}$
is approached from above, this fact together with the coupling between
the two sectors leads to a dispersion
%\begin{equation}
$\epsilon_1(k) \approx ({v_F^2}k^2/{4\Delta}) - \Delta$
%\end{equation}
of the ``incommensurate'' soft mode with $\Delta \propto
\left(H-H_{c2}\right)^2$ rather than the usual behaviour $\Delta
\propto \left|H-H_{c2}\right|$ for fermions.  This field dependence
immediately gives the linear $H$-dependence of the magnetization for
$H\gtrsim H_{c2}$.
%%%%%%%%%%%%%%%%%%%%%%%%%%%%%%%%%%%%%%%%%%%%%%%%%%%%%%%%%%%%%%%%%%%%%%

In summary we have studied the properties of doped Heisenberg chains
in a magnetic field in the framework of a Bethe Ansatz solvable model.
We find plateaus in the magnetization curves at certain values $M_p$
of the magnetization.  To our knowledge this is a novel feature in an
integrable model thereby providing the basis for more detailed studies
aiming at a better understanding of the mechanism for the occurence of
these plateaus and the critical behaviour at their end points.  $M_p$
can be continuously tuned by changing the concentration of carriers.
This is a fundamentally different from the fixed values of $M_p$
obtained from topological arguments in the magnetic insulators studied
previously.
An extension of such arguments to the case of doped chains might be
possible within a \emph{classical} treatment of the localized $t_{2g}$
spins in the double-exchange model: in this limit the ground state has
an incommensurate magnetic structure with $k_F$ periodicity
\cite{Koshi99}.  However, unlike the models considered here this
approach also leads to gapped charge excitations.
An alternative attempt to describe the plateaus at tunable fractions
of the maximal magnetization within a bosonized model relies just on
this existence of a gapless charge mode \cite{cabra:pc}.
Similarly, the second feature where the plateaus discussed here differ
from the ones in quantum spin chains and ladders can be understood
only as a consequence of the existence of a second massless mode: the
relation of the physical quantum numbers to the conserved charges of
the effective low energy theory together with the coupling of the
gapless channels conspire to give the linear field dependence of the
singular part of the magnetization observed near the critical point
$H_{c2}$.  Note that this mechanism does not restrict the critical
exponents to the values $1/2$ and $1$ discussed in this letter.

Further studies of the low energy properties --- in particular the
analysis of the asymptotics of correlation functions as $H_{c1,2}$ are
approached --- in this solvable microscopic model will lead to new
insights into the critical behaviour at the plateau transitions and
possible the related ones into Mott insulating phases of interacting
particles.
Furthermore, the phenomena reported in this letter may be verified in
experimental studies of the magnetization process in the doped,
effectively one-dimensional transition metal oxides mentioned above.

%\section*{Acknowledgements}
We thank D.\,C.\ Cabra, F.\,H.\,L.\ E\ss{}ler and A.\,M.\ Tsvelik for
discussions.  This work is supported in parts by the Deutsche
For\-schungs\-gemeinschaft under Grant No.\ Fr~737/2.

%\bibliographystyle{h-physrev}
%\bibliography{base,plat,corr,bound,books,frahm}

\narrowtext

\end{multicols}

\end{document}